\documentclass[journal=mamobx,manuscript=article]{achemso}   
\pdfoutput=1 
\usepackage{graphicx}%

\usepackage[intlimits]{amsmath}
\usepackage{amsfonts}
\usepackage{amsthm}
\usepackage{mathrsfs}
\usepackage{stmaryrd}

\newcommand{\ie}{\emph{i.e.},~}

\newcommand{\etal}{\emph{et al.~}}

\title[Optimizing ELFSE]{Optimizing end-labeled free-solution electrophoresis by
increasing the hydrodynamic friction of the drag-tag}
\author{Kai Grass}
\email{grass@fias.uni-frankfurt.de}
\author{Christian Holm}
\email{holm@icp.uni-stuttgart.de}
\affiliation{Frankfurt Institute for Advanced Studies, Goethe
University, Ruth-Moufang-Strasse 1, D-60438 Frankfurt am Main, Germany}
\alsoaffiliation{Institute for Computational Physics, University of
Stuttgart, Pfaffenwaldring 27, D-70569 Stuttgart, Germany}
\author{Gary W. Slater}
\email{gslater@uottawa.ca}
\affiliation{Department of Physics, University of Ottawa, 150 Louis-Pasteur,
Ottawa, Ontario K1N 6N5, Canada}
 
\begin{document}

\section*{Abstract}
 
We study the electrophoretic separation of polyelectrolytes of varying lengths by
means of end-labeled free-solution electrophoresis (ELFSE). A coarse-grained
molecular dynamics simulation model, using full electrostatic interactions and a
mesoscopic Lattice Boltzmann fluid to account for hydrodynamic interactions, is
used to characterize the drag coefficients of different label types: linear and
branched polymeric labels, as well as transiently bound micelles. 

It is specifically shown that the label's drag coefficient is determined by its
hydrodynamic size, and that the drag per label monomer is largest for linear
labels. However,  the addition of side chains to a linear label offers the
possibility to increase the hydrodynamic size, and therefore the label
efficiency, without having to increase the linear length of the label, thereby
simplifying synthesis. The third class of labels investigated, transiently bound
micelles, seems very promising for the usage in ELFSE, as they provide a
significant higher hydrodynamic drag than the other label types.

The results are compared to theoretical predictions, and we investigate how the
efficiency of the ELFSE method  can be improved by using smartly designed
drag-tags.

\section{Introduction}

As known from experiments and
theory~\cite{stellwagen03a,hoagland99a,cottet00b,muthukumar96a,volkel95b,mohanty99a},
the free-solution mobility of a flexible polyelectrolyte chain does not depend on
the chain length $N$ (number of monomers) anymore if the chain is longer than a
certain length $N_\text{FD}$. The regime where $N > N_\text{FD}$ is called free-draining regime.
In this regime, the counterions influence the inter-monomer hydrodynamic
interactions and allow the fluid to drain through the polyelectrolyte coil. The
effective friction $\Gamma_\text{eff}$ becomes linear in the chain length, as
does the effective charge $Q_\text{eff}$ for longer chains, which leads to a
constant, length-independent mobility
\begin{equation}
	\mu_0 = \frac{Q\text{eff}}{\Gamma_\text{eff}}.
\end{equation}

It was shown that attaching a suitable uncharged molecule to an electrophoresis
target can restore the size-dependent mobility and overcome the free-draining
property of long polyelectrolyte chains~\cite{heller98a,ren99a,desruisseaux01a}.
This method, which is known as \emph{end-labeled free-solution electrophoresis}
(ELFSE), is based on the alteration of the charge-to-friction ratio of the
polyelectrolyte molecules by an uncharged drag label.

The effect of the label can be compared to that of a parachute attached to a
moving object. The additional friction provided by the parachute slows the object
down. This effect is stronger for smaller molecules with a lower effective
friction and smaller charge to pull the drag-tag, as the ratio between charge
and friction is changed more drastically than for larger molecules.

Since the method's introduction, finding suitable labels that provide a high
hydrodynamic drag has been a major concern in this
field~\cite{meagher05a,meagher06a,mccormick05a}. A larger hydrodynamic drag
enables the separation of longer chain fragments, as the length-dependence of the
electrophoretic mobility decreases with increasing polyelectrolyte length. When
the additional friction provided by the drag-tag becomes negligible against the
intrinsic effective friction of the polyelectrolyte, the chain becomes
essentially free-draining again. Experimentally relevant is that the mobility of
long polyelectrolyte chains should differ by a factor large enough to allow for
accurately separating them, although their lengths only vary by a single
monomer. The maximum chain length resolvable in this way is called the
\emph{read length}.

In general, the drag labels can be chosen from a wide range of molecules but they
have to fulfill certain requirements, such as being water-soluble at experimental
conditions, having a unique attachment mechanism to the polyelectrolyte and
showing minimal interaction with the walls of the capillary. The read length is
optimised by choosing a large molecule that imposes a high frictional drag.
However, to fulfill resolution requirements, the labels must remain perfectly
monodisperse since polydispersity will effectively be like an additional source
of diffusion that would broaden the peaks.

As it poses an experimental challenge to produce large, monodisperse linear
polymer labels, two recently proposed alternatives seem promising.
Haynes~et.~al.~\cite{haynes05a} proposed to use branched polymers with
well-defined architectures. A first theoretical study on this
method~\cite{nedelcu05a} verified the approach and concluded that, even though a
branched polymer is more compact and thus provides a smaller hydrodynamic
friction for a given molecular weight than a linear polymer, this drawback is
offset by the monodispersity of the branched labels created by assembling shorter
linear chains. Grosser~et.~al.~\cite{grosser07a,savard07a,savard08a} introduced
nonionic surfactant micelles as drag labels with very large hydrodynamic
friction. The inherent polydispersity of the micelles is overcome by using a PNA
amphiphile that only provides a transient binding between the DNA fragments to be
separated and the micelles. Each fragment attaches to a different micelle every
couple of seconds, which results in an averaging procedure over the course of the
elution time that remedies the need for perfect monodispersity. 

The theoretical description of the methods discussed above is not complete and
furthermore predicts behaviours that are difficult to test experimentally, such
as end of chain effects, the hydrodynamic deformation of the label in high
fields, or the steric segregation between the label and the chain. Since it is
not possible to visualize DNA-label molecules in the lab, computer simulations
can support the understanding of the real physics as long as they include
hydrodynamic interactions between polyelectrolyte, label and solvent, as well as
account for the influence of the electrostatic interaction between the
polyelectrolytes and its surrounding counterions. Of course, it is beyond the
scope of this article to cover all previous
predictions~\cite{mccormick01a,mccormick05a,mccormick07a,mccormick07b,nedelcu05a,nedelcu07a}.
However, we will demonstrate that it is possible to study these factors, and that
the standard theory appears to be sufficient for the cases treated here. More
cases will be studied in future papers.

Since the ELFSE method overcomes the main drawback of ordinary gel
electrophoresis, the long separation time due to the slow down by the applied 
gel matrix, it is a promising method on the
way to faster sequencing methods and, as such, of especial interest to the
community. 

In this paper we will use coarse-grained MD simulations to study the
electrophoretic separation of fully flexible polyelectrolytes of varying lengths
by end-labeling. After introducing the simulation model, we confirm that the
free-draining behaviour is correctly reproduced and test the standard theory for
ELFSE by attaching an uncharged linear label. In Section~\ref{sec:branchedlabel},
we introduce branches to the drag label and test the predictions made
by~\cite{nedelcu05a,haynes05a}. From the branched label, we will go to micellar
targets (Section~\ref{sec:micellelabel}) and analyze the method proposed
by~\cite{savard07a,grosser07a}. We establish a relation between the average size
of the micelle and its drag value. Our concluding remarks point out the
efficiency of the ELFSE method and show the benefit of the different labeling
methods.

\section{Theory}

The theory for end-labeled free-solution electrophoresis is based on the
interplay between hydrodynamic and electrostatic forces, and it takes into
account the stress that builds along the chain backbone. In general, it is
assumed that the electrostatic and frictional forces do not deform the hybrid
molecule's random coil conformation nor its cloud of counterions. Because of
these assumptions, the theory used here is valid for low velocities and weak
electric fields.

Neglecting molecular end-effects, the electrophoretic mobility $\mu = v/E$ of
the polyelectrolyte with an attached linear drag-tag can be described in terms of
the effective friction of the polyelectrolyte $\Gamma_\text{PE}$, its
effective charge $Q_\text{PE}$ and the hydrodynamic friction of the attached
label $\Gamma_\text{L}$:
\begin{equation}\label{eq:elfse-mobility}
	\mu = \frac{Q_\text{PE}}{\Gamma_\text{PE}+\Gamma_\text{L}} = \mu_0
	\frac{1}{1+\Gamma_\text{L}/\Gamma_\text{PE}},
\end{equation}
where $\mu_0$ is the length independent free solution mobility without
drag-tag. 

Equation~\ref{eq:elfse-mobility} shows the importance of the ratio between
$\Gamma_\text{PE}$ and $\Gamma_\text{L}$. The electrophoretic mobility $\mu$ is
a function of $N$ for a fixed $\Gamma_\text{L}$ as long as $\Gamma_\text{PE}$
changes with $N$ and the ratio between $\Gamma_\text{L}$ and $\Gamma_\text{PE}$
remains non-negligible.  

Since the electrophoretic friction coefficient $\Gamma_\text{PE}$ grows linear
with the length of the polyelectrolyte for long chains, as shown in a previous
publication~\cite{grass08c}, Equation~\ref{eq:elfse-mobility} can be reformulated
as follows:
\begin{equation}\label{eq:elfse-alpha}
	\mu = \mu_0 
	\frac{1}{1+\alpha_\text{L}/N},
\end{equation}
with a constant drag coefficient
\begin{equation}
	\alpha_\text{L} = \frac{\Gamma_\text{L}}{\Gamma_\text{PE}/N}.
\end{equation}
Here the ratio $\Gamma_\text{PE}/N$ is the friction per monomer of the
polyelectrolyte. The chemistry and temperature dependent $\alpha_\text{L}$ is a
measure for the difference in hydrodynamic properties between the polyelectrolyte and the
label and represents the number of polyelectrolyte monomers that provide a
hydrodynamic friction equal to that of the
label~\cite{ren99a,long98a,mccormick01a,mccormick05a,mccormick07a}.

In order to characterize the effectiveness of an arbitrary (not necessarily
linear) label for size-separation, Equation~\ref{eq:elfse-alpha} has been used
to define this specific label property from the measured mobilities:
\begin{equation}
	\alpha_\text{L} = N \left(\frac{\mu_0}{\mu} -
	1\right),
\end{equation}
where $\alpha_\text{L}$ is conveniently determined as the slope when plotting
$\mu_0/\mu$ versus $1/N_\text{PE}$:
\begin{equation}\label{eq:elfse-alphal}
	\frac{\mu_0}{\mu} = 1 + \alpha_L/N.
\end{equation}

\section{Simulation model}

We employ coarse-grained molecular dynamics (MD) simulations using the ESPResSo
package~\cite{limbach06a} to study the electrophoretic separation of fully
charged linear flexible polyelectrolytes by end-labeled free-solution
electrophoresis. The polyelectrolytes are modelled by a totally flexible bead-spring model as a set of
spheres that represent the $N$ individual monomers which are connected to each
other by finitely extensible nonlinear elastic (FENE) bonds~\cite{soddeman01a}:
\[
U_\text{FENE}(r) = \frac{1}{2} k R^2 \ln \left( 1 - \left( \frac{r}{R}
\right)^2 \right),
\]
with stiffness $k = 30 \epsilon_0$, and maximum extension $R = 1.5 \sigma_0$, and $r$
being the distance between the interacting monomers.
Additionally, a truncated Lennard-Jones or WCA potential is used for excluded
volume interactions~\cite{weeks71a}:
\[
U_\text{LJ}(r) = \epsilon_0 \left( \left( \frac{\sigma}{r}\right)^{12} -
\left( \frac{\sigma}{r}\right)^6 + \frac{1}{4} \right),
\]
with the cut-off being $r_\text{cut} = \sqrt[6]{2}\sigma$, at which
$U_\text{LJ}(r) = 0$.

The parameter $\epsilon_0$ and $\sigma_0$ define the energy and length scale of
the simulations. We use $\epsilon_0 = k_\text{B} T$, i.e.~the energy of the
system is expressed in terms of the thermal energy. The length scale $\sigma_0$
defines the size of the monomers. We set $\sigma_0 = 4.7 \text{\AA}$, which with
the average bond length along the PE chain being $0.91 \sigma_0$ represents a
linear monomer distance of approximately 4.3 \AA, the spacing of
single-stranded DNA. Different polyelectrolytes can be mapped by changing
$\sigma_0$. Unless mentioned otherwise, all observables are expressed in terms of
the simulation units $\sigma_0$ and $\epsilon_0$, which we will not use
explicitly from now on.

Besides the dissociated counterions the system also contains additional
monovalent salt. The counterions and the salt ions are modelled as charged
spheres using the same WCA potential giving all particles in the system the
same size.

All chain monomers carry a negative electric charge $q
= -1 e_0$, where $e_0$ is the elementary charge. For charge neutrality, $N$ monovalent
counterions of charge $+1 e_0$ are added. Additional
monovalent salt is added to the simulation. Full electrostatic interactions are
calculated with the P3M algorithm using the implementation of
Reference ~\cite{deserno98a}. The Bjerrum length $l_B = e_0^2 / 4 \pi \epsilon
k_\text{B} T = 1.5$ in simulation units corresponds to 7.1 \AA, the Bjerrum
length in water at room temperature. This means that the effect of the surrounding water is modelled implicitly by
simply employing the dielectric properties of water, using a relative
dielectric constant of $\epsilon_r \approx 80$. The applied external field $E
= 0.1$ is represented by a constant force, acting on all charged particles. The
field strength is chosen to be sufficiently small in order to not influence
chain conformations or couterion distributions.

\begin{figure}[htp]
\begin{center}
\includegraphics[width=\textwidth]{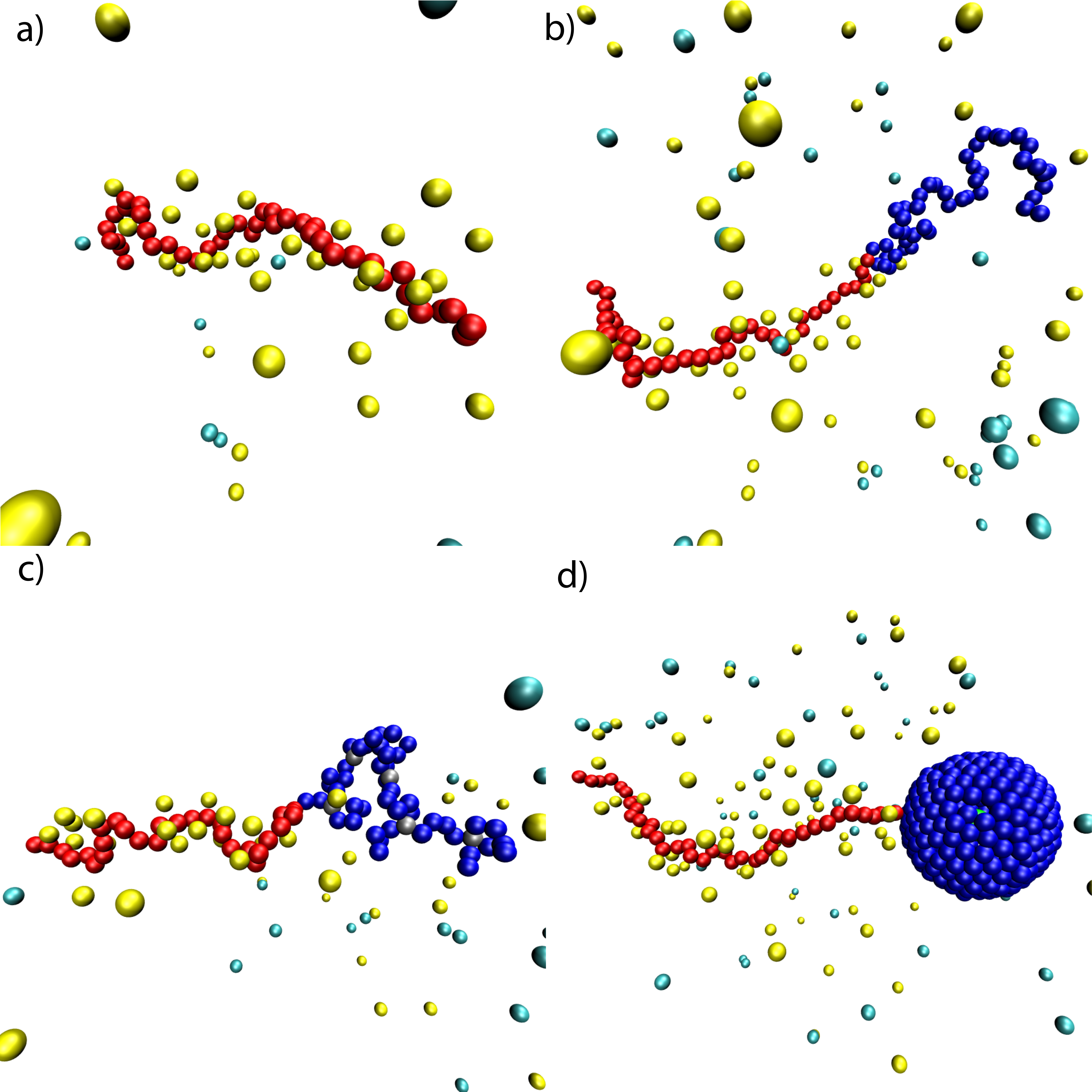}
\caption{(a) Polyelectrolyte with surrounding counter- and co-ions. (b)
  with linear drag-tag, (c) with branched polymeric drag-tag, and (d) with micellar 
 drag-tag.}
\label{fig:elfse-dragtags}
\end{center}
\end{figure}

The coarse-grained molecular dynamics model used in this article is extended by 
the inclusion of the three drag-tag labels investigated, as shown in
Figure~\ref{fig:elfse-dragtags}. The linear labels use the same flexible
bead-spring model as the polyelectrolytes, but are uncharged. For the study of
branched labels, flexible side chains of well-defined length are added to the
linear label. The third kind of label, micellar drag-tags, is represented by a
sphere of given radius whose surface is modelled by many small WCA spheres that
are connected with each other by a network of FENE springs. The number of small
spheres is defined by the radius of the large sphere to be modelled. This model
has been successfully used to study colloidal
electrophoresis~\cite{lobaskin04a,lobaskin04b,lobaskin07a}.

We include hydrodynamics using a Lattice Boltzmann (LB)
algorithm~\cite{mcnamara88a} that is frictionally coupled to the MD simulations
via an algorithm introduced by Ahlrichs et al.~\cite{ahlrichs99a}. The mesoscopic
LB fluid is described by a velocity field generated by discrete momentum
distributions on a spatial grid, rather than explicit fluid particles. We use an
implementation of the D3Q19 model with a kinematic viscosity $\nu = 1.0$, a fluid
density $\rho = 1.0$~\cite{duenweg07a}. The resulting fluid has a dynamic
viscosity $\eta = \rho \nu = 1.0$. The space is discretized by a grid with
spacing $a = 1.0$. The fluid is coupled to the particles by a frictional coupling
with bare friction parameter $\Gamma_\text{bare} = 20.0$. Random fluctuations for
particles and fluid act as a thermostat. The interaction between particles and
fluid conserve total momentum and are proved to yield correct long-range
hydrodynamic interaction between individual particles.

The simulations are carried out under periodic boundary conditions in a
rectangular simulation box. We investigate the behaviour of polyelectrolyte
chains varying from $N=20$ to $N=60$ monomers. The size $L$ of the box is
varied to realize a constant monomer density of $n_\text{PE} = 10^{-3}$
which corresponds to a concentration $c_\text{PE} = 16 \text{mM}$. The same
concentration is used for the additional salt, resulting in a Debye length of
$\lambda_\text{D} \approx 4.2$.  A MD time step $\tau_\mathrm{MD}
= 0.01$ and a LB time step $\tau_\text{LB} = 0.05$ are used. After equilibration of
$10^6$ steps, $10^7$ steps are used for generating the data. The time-series are analyzed
using auto-correlation functions to estimate the statistical errors as detailed
in Reference~\cite{wolff04a}. Error bars of the order of the symbol size or smaller
are omitted in the figures. 

The electrophoretic mobility is obtained by applying a constant electric field of
reduced field strength $E=0.1$ that acts on all charged particles. The mobility
is then given by direct measurement of the center of mass velocity $v$ of the
chain:
\begin{equation}
	\mu = \frac{v}{E}.
\end{equation}
Before applying this method, it was ensured that the applied electric field
strength $E$ is small enough not to distort chain conformations or counterion
distributions. Therefore, the system is in the linear
response regime, \ie the measured mobility does not depend on the magnitude of
the electric field.

Up to ten independent simulations are carried out for each data point, taking
between one day and two weeks on a single standard CPU\footnote{Dual Core AMD
Opteron(tm) Processor 270} depending on the chain length $N$ and the type of
label investigated.

\section{Linear drag tags}

In this section, the simulation model is applied to the electrophoresis of
polyelectrolyte chains with an attached linear polymeric drag-tag. The
electrophoretic mobility for polyelectrolyte chains is determined with and
without different labels, and the results are compared to the theoretical
predictions. We also examine how the effective friction of the drag-tag is
influenced by the intrinsic stiffness and the salt concentration in the
solution.

\subsection{Testing the standard ELFSE theory}

\begin{figure}[htp]
\begin{center}
  \includegraphics[width=\textwidth]{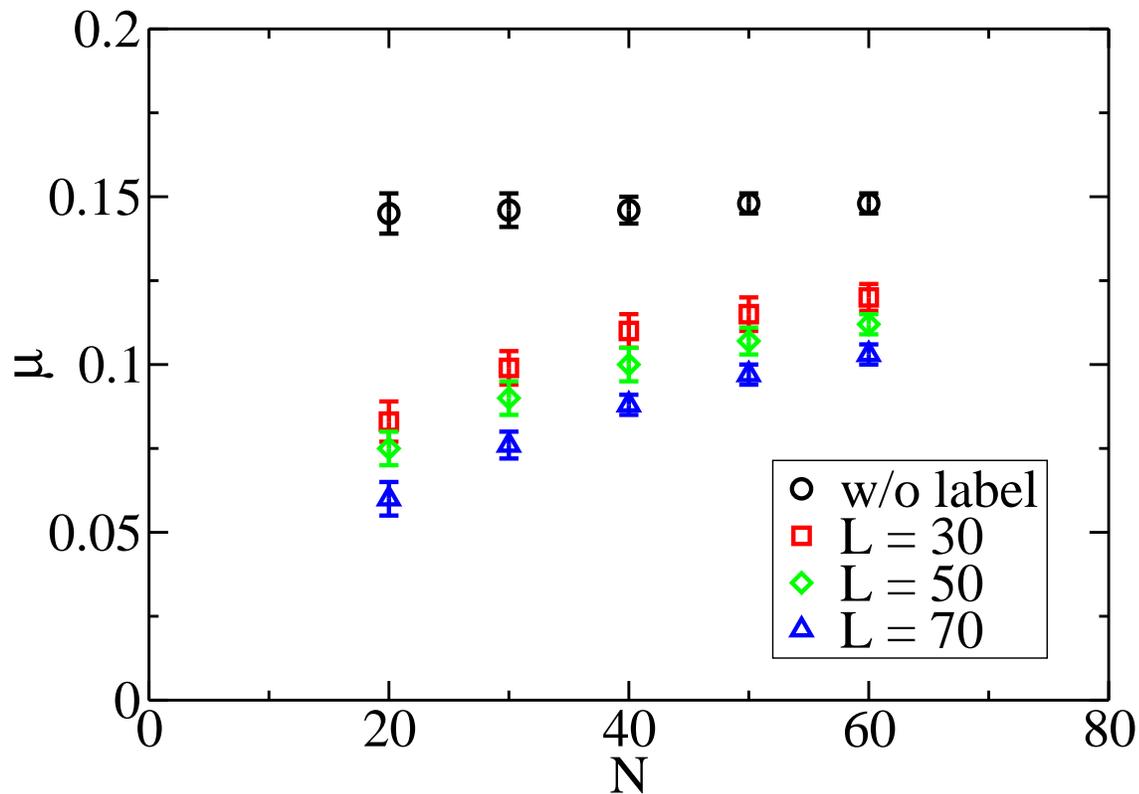}
  \caption{The free-solution electrophoretic mobility without label (black
  circles) shows no dependence on the chain length $N$. The free-draining
  mobility is $\mu_0 = 0.147 \pm 0.002$. The attachment of linear drag-tags to
  the end of the polyelectrolyte chains reduces the mobility and restore a
  $N$-dependent behaviour. The label length $L$ is varied from 30 to 70
  monomers, with the largest label resulting in the strongest slowdown.}
  \label{fig:elfse-linearlabel}
\end{center}
\end{figure}

First, the free-solution electrophoretic mobility without an attached drag-tag,
$\mu_0$, is determined, as shown in Figure~\ref{fig:elfse-linearlabel}. The
measured mobility does not depend on the chain length, as expected for longer
free-draining polyelectrolyte chains. The average mobility is determined to be
\begin{equation}
	\mu_0 = 0.147 \pm 0.002.
\end{equation}

\begin{figure}[htp]
\begin{center}
  \includegraphics[width=\textwidth]{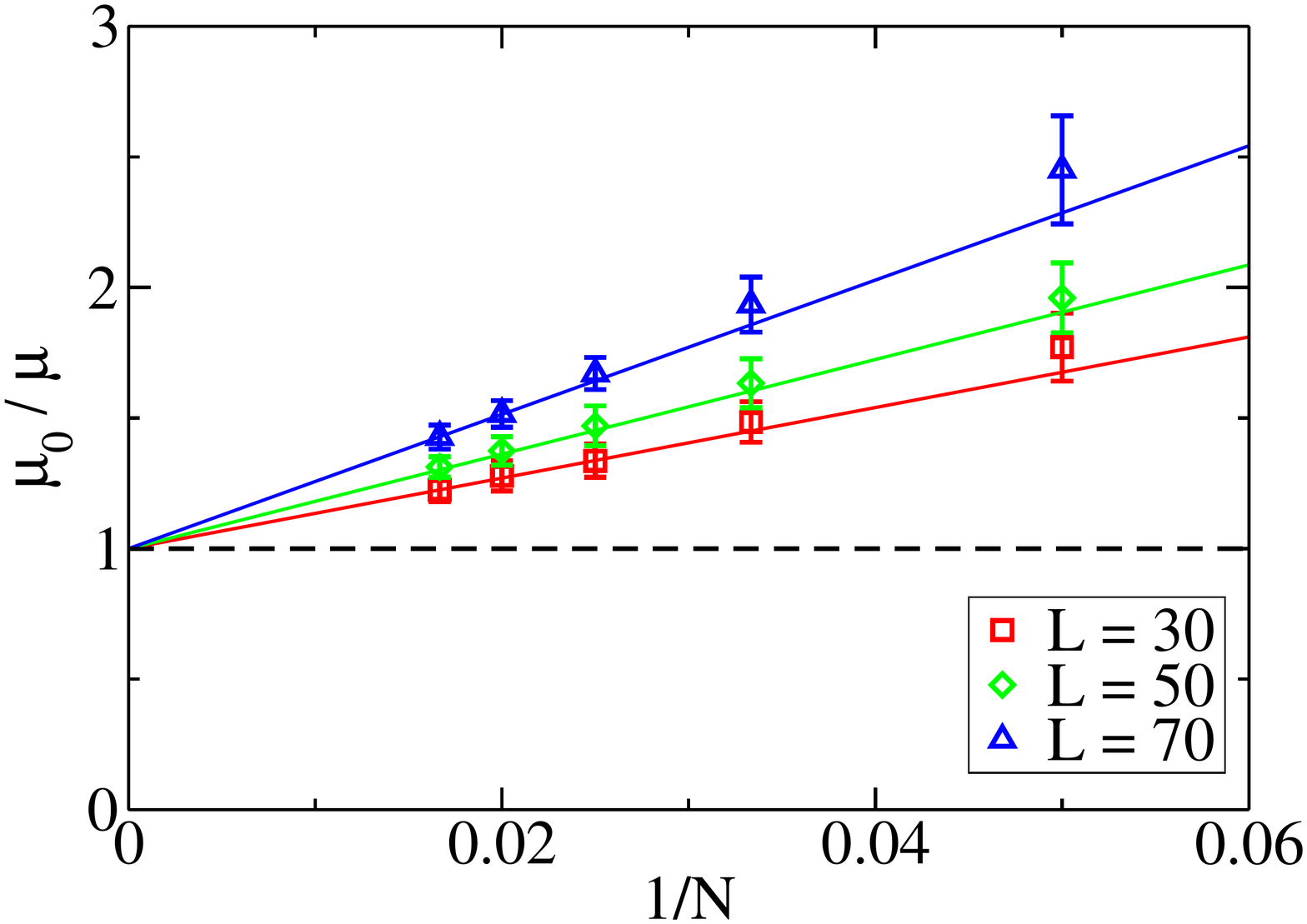}
  \caption{The hydrodynamic drag coefficient
  $\alpha_\text{L}$ is given by the slope of the curve. For the linear labels,
  $\alpha_\text{L}$ ranges from $13.5 \pm 0.4$ to $25.7 \pm 0.6$.}
  \label{fig:elfse-linearalpha}
\end{center}
\end{figure}

Additionally, in Figure~\ref{fig:elfse-linearlabel}, the mobilities with attached
drag-tags ranging from $L=30$ to $L=70$ monomers are measured, and it is confirmed
that a length-dependence is achieved and that the difference in mobilities, \ie
the selectivity of the separation, is better the longer the attached label
is. Equation~\ref{eq:elfse-alphal} is used to calculate the hydrodynamic drag
coefficients as shown in Figure~\ref{fig:elfse-linearalpha}, resulting in values
from $\alpha_\text{L} = 13.5 \pm 0.4$ for $L=30$ to $\alpha_\text{L} = 25.7 \pm
0.6$ for $L=70$.

In the following, an expression for $\alpha_\text{L}$ based on the hydrodynamic
size and shape of the label is developed. The hydrodynamic friction
$\Gamma_\text{L}$ of the uncharged label is related to the hydrodynamic radius
$R_\text{h}$ by means of the Stokes relation:
\begin{equation}
	\Gamma_\text{L} = 6 \pi \eta R_\text{h,L}.
\end{equation}
As in the previous chapter, the effective electrophoretic friction of the
polyelectrolyte is expressed in terms of the free-solution mobility $\mu_0$ and
the effective charge $Q_\text{eff}$:
\begin{equation}
	\Gamma_\text{PE} = Q_\text{eff} / \mu_0.
\end{equation}
Using the Manning prediction $Q_\text{eff} = \left(1/\xi\right) N$ for the
effective charge~\cite{manning98a}, where the Manning condensation parameter
$\xi = l_\text{B}/b$ is a measure for the strength of the electrostatic potential of the
polyelectrolyte, finally yields
\begin{equation}\label{eq:elfse-alpharh}
	\alpha_\text{L} = \mu_0 \xi 6 \pi \eta R_\text{h,L}.
\end{equation}
With the system parameter used here, $\xi = 1.63$, one obtains
\begin{equation}
	\alpha_\text{L} = \left( 4.5 \pm 0.1 \right) R_\text{h,L}. 
\end{equation}

Equation~\ref{eq:elfse-alpharh} will be shown to be valid for linear labels whose
size is not exceeding the Debye length $\lambda_\text{D}$. When the label size
becomes larger, the friction of the label is not anymore directly related
to the hydrodynamic radius, as the salt ions that penetrate the polymer coil
influence the intermonomer hydrodynamic interactions and limit them to the
electrostatic screening length. As for the polyelectrolyte itself, this
screening length is of the order of the Debye length.

For linear labels larger than the Debye length, McCormick \etal introduced a
relation for the hydrodynamic drag coefficient, with which $\alpha_\text{L}$ can
be determined from the size of the
polyelectrolyte and label monomers, $b_\text{PE}$ and $b_\text{L}$, and the
corresponding Kuhn lengths, $b_\text{k,PE}$ and $b_\text{k,L}$, which describe
the stiffness of the chains~\cite{mccormick01a}:
\begin{equation}\label{eq:elfse-alphamccormick}
	\alpha_\text{L} = \frac{b_\text{L} b_\text{k,L}}{b_\text{PE} b_\text{k,PE}} L.
\end{equation}

\begin{figure}[htp]
\begin{center}
  \includegraphics[width=\textwidth]{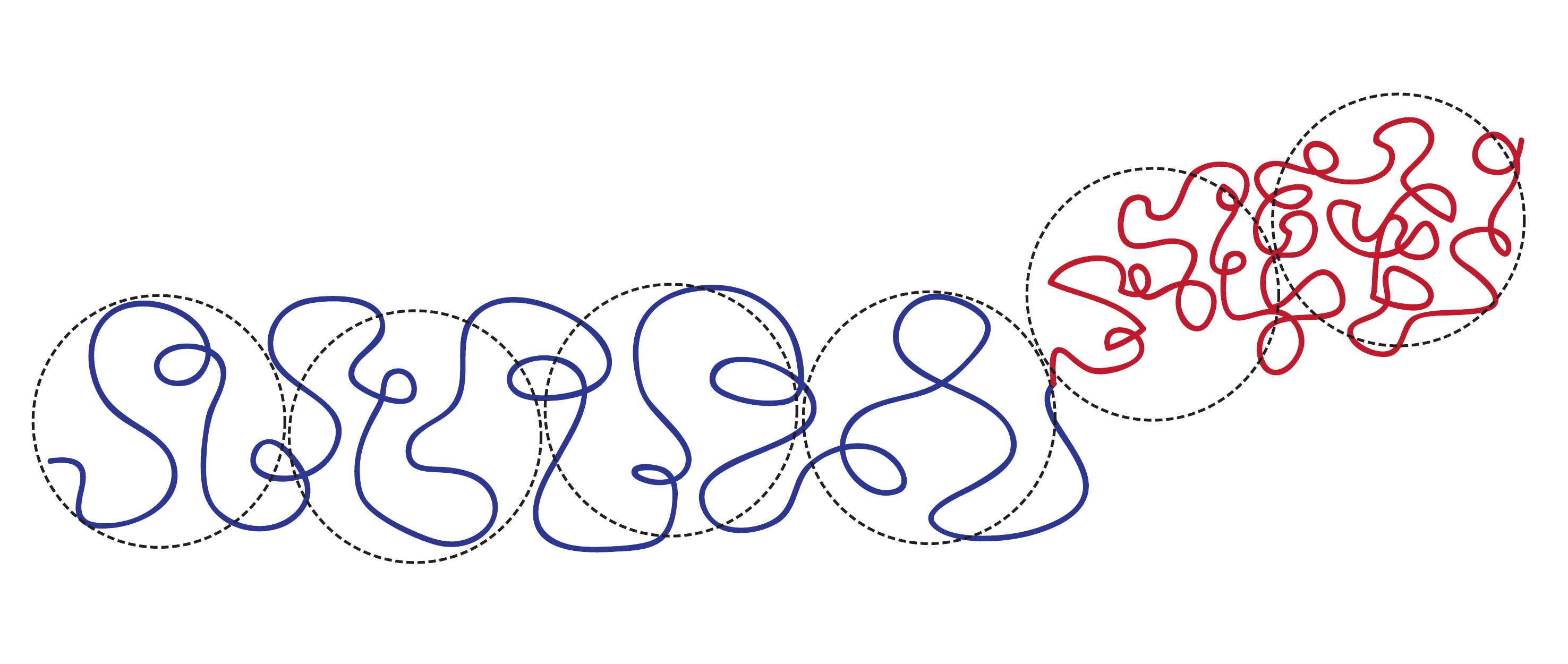}
  \caption{A schematic representation of the ``blob'' picture used to derive
  Equation~\ref{eq:elfse-alphamccormick}.}
  \label{fig:elfse-blobs}
\end{center}
\end{figure}
	
The derivation of Equation~\ref{eq:elfse-alphamccormick} assumes that the
polyelectrolyte and the label can be represented by a series of
hydrodynamically equivalent entities, called ``blobs'' as shown in
Figure~\ref{fig:elfse-blobs}. The number and the size of these blobs depend on
the bond length and flexibility of the chains, resulting in the presented
relation for $\alpha_\text{L}$.

The total effective friction of the polyelectrolyte-label compound with the
surrounding solvent is linear in the total number of hydrodynamically equivalent
monomers given by
\begin{equation}
	N = N_\text{PE} + \alpha N_\text{L}. 
\end{equation}
This is true for long polyelectrolytes in the free-draining regime, where the
size of the compound is larger than the Debye length $\lambda_D$, since the
hydrodynamic interactions between the individual monomers are screened on this
length scale, as shown in a prior study on
free-solution electrophoresis~\cite{grass08a,grass08c}.

Thus, the hydrodynamic drag $\alpha_\text{L}$ can be directly
calculated from the persistence lengths of the polyelectrolyte and of the label using
Equation~\ref{eq:elfse-alphamccormick}. Here, $l_\text{p,PE}$ and $l_\text{p,L}$
are calculated from the bond correlation function~\cite{hsiao06b}:
\begin{equation}\label{eq:elfse-persistence}
		l_p = \frac{1}{2 b} \sum_{i=0}^{N/2} \langle
		\vec{b}_{{N/2}}\cdot\vec{b}_{{N/2+i}} + \vec{b}_{{N/2}}\cdot\vec{b}_{{N/2-i}} \rangle,
\end{equation}
where $\vec{b}_{i}$ is the $i$-th bond vector and $b$ is the average bond length.
The angular brackets $\left< \ldots \right>$ denote an ensemble
average.\footnote{For a discussion about different ways to determine the
persistence length in computer simulations please refer to~\cite{ullner02a}.}

Under the chosen conditions, the persistence length of the polyelectrolyte is
found to be
\begin{displaymath}
 l_\text{p,PE} = 5.1 \pm 0.3,
\end{displaymath} 
and the label's one
\begin{displaymath}
 l_\text{p,L} = 1.9 \pm 0.1.
\end{displaymath}
The difference between these two values is due to the electrostatic repulsion
between the monomers of the polyelectrolyte. In our model, all monomers have the
same size, so that Equation~\ref{eq:elfse-alphamccormick} is reduced to
\begin{equation}\label{eq:alphalinear}
	\alpha_\text{L} = \frac{l_\text{p,L}}{l_\text{p,PE}} L = \left( 0.37 \pm
	0.03\right) L.
\end{equation}

\begin{figure}[htp]
\begin{center}
  \includegraphics[width=\textwidth]{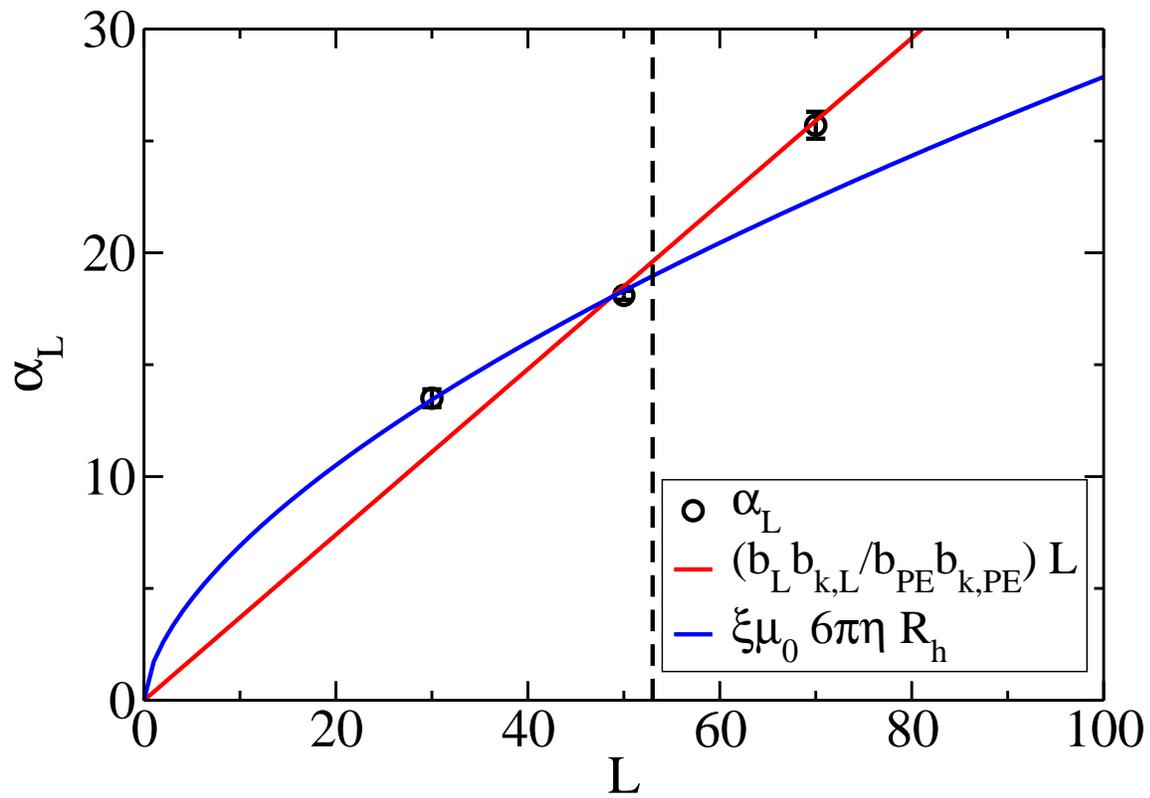}
  \caption{The hydrodynamic drag coefficient $\alpha_\text{L}$ for the linear
  label is compared to the theoretical predictions of
  Equations~\ref{eq:elfse-alpharh} and~\ref{eq:elfse-alphamccormick}. Both
  predictions are valid in different regimes with the division being
  $R_\text{h} \approx \lambda_\text{D}$ indicated by the vertical dashed line.
  Note that neither theory has a free fitting parameter used to achieve the
  quantitative agreement with the simulation data. }
  \label{fig:elfse-linearalphacomparison}
\end{center}
\end{figure}

The comparison between the measured drag coefficient and the theoretical
prediction in Figure~\ref{fig:elfse-linearalphacomparison} shows an agreement for
the respective regimes of validity. The agreement between theory and experiments
has been shown before~\cite{vreeland01a,mccormick01a,meagher05a}. For labels with
a hydrodynamic size smaller than the Debye length, \ie $R_\text{h} <
\lambda_\text{D}$, Equation~\ref{eq:elfse-alpharh} gives the correct prediction
for the drag coefficient $\alpha_\text{L}$. Longer labels, however, can no longer
be seen as a single polymer coil with a hydrodynamic size $R_\text{h}$, but
instead the blob picture described by Equation~\ref{eq:elfse-alphamccormick} has
to be used. This prediction is only valid when the hydrodynamic size becomes
larger than the Debye length. For $R_\text{h} \approx \lambda_\text{D}$ the
expected cross-over between these regimes is observed.

It remains to be emphasised that, by
determining $\alpha_\text{L}$ from the measurements of the persistence lengths
and the hydrodynamic radius, there is \emph{no free} fitting parameter and the
\emph{quantitative} agreement in
Figure~\ref{fig:elfse-linearalphacomparison} is noteworthy.

\subsection{Increasing the hydrodynamic drag coefficient}

In Figure~\ref{fig:elfse-linearalphacomparison}, it is shown that the total drag
coefficient $\alpha_\text{L}$ for linear labels can be increased by using longer
labels, and that beyond the Debye length the increase is linear with the length
$L$ of the label. Unfortunately, the experimental requirement of strict
monodispersity of the label limits the size of linear polymeric labels that can
be synthesized and prepared. In this section, it will be shown how one can
influence the total drag coefficient also by modifying the relative stiffness of
the polyelectrolyte and drag-tag chains.

\subsubsection{Increasing the label stiffness}

Equation~\ref{eq:elfse-alphamccormick} shows the dependence of $\alpha_\text{L}$ on the persistence lengths
of polyelectrolyte and label. Therefore, $\alpha_\text{L}$ can be increased by either
increasing the persistence of the label or decreasing the persistence of the
polyelectrolyte. Both ways will be investigated in this subsection.

First, an additional harmonic bond angle potential,
\begin{equation}
	U_\text{BA} = k_\text{BA}\left(\phi-\phi_0\right)^2,
\end{equation}
is added to the interaction between the label monomers, where $\phi$ is the angle
between two consecutive bonds. Here, $k_\text{BA} = 30$ and $\phi_0 =
0$ are chosen.

The bond angle potential increases the hydrodynamic
radius of the 30 monomer label from $R_\text{h,L} = 3.00 \pm 0.05$ to
\begin{displaymath}
R_\text{h,L} = 5.25 \pm 0.05,
\end{displaymath}
and thus puts the label size into the regime where the blob picture is valid.
The increased stiffness doubles the persistence length of the label from
$l_\text{p,L} = 1.9 \pm 0.1$ to
\begin{displaymath}
 l_\text{p,L} = 4.0 \pm 0.1,
\end{displaymath}
which yields an increased drag coefficient according to
Equation~\ref{eq:elfse-alphamccormick} of:
\begin{displaymath}
\alpha_\text{L} = \left( 0.79 \pm 0.04 \right) L \approx 23.7. 
\end{displaymath}

\begin{figure}[htp]
\begin{center}
  \includegraphics[width=\textwidth]{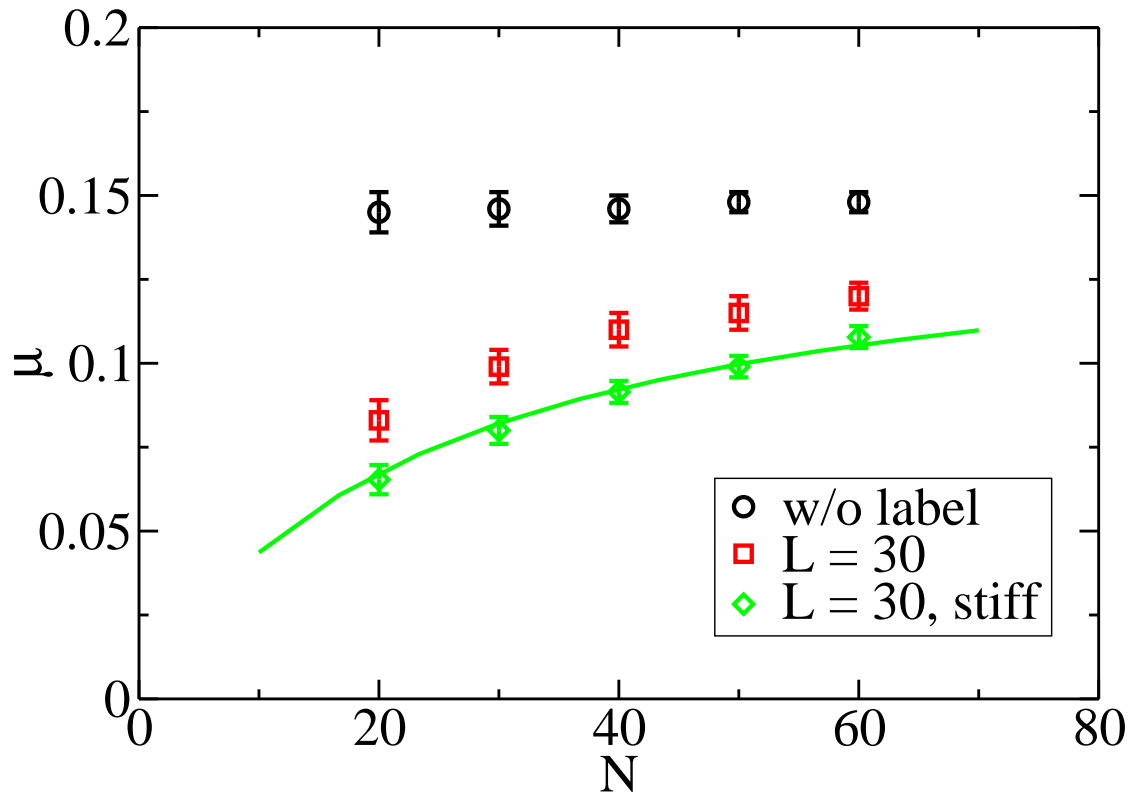}
  \caption{The hydrodynamic drag coefficient of a stiff linear label is
  higher than that of a fully flexible label with the same length. The
  slowdown of the stiff label is correctly predicted by
  Equation~\ref{eq:elfse-alphamccormick} (solid line).}
  \label{fig:elfse-stifflabel}
\end{center}
\end{figure} 

Figure~\ref{fig:elfse-stifflabel} compares the theoretical predicted slowdown of
the stiffer 30 mono\-mer\-ic label to the measured mobilities. As before, an
excellent agreement to the theory is found for the investigated label lengths.
Please note that there is no fitting parameter here.

\subsubsection{Reducing the polyelectrolyte stiffness}

In practice, it is not easy to increase the stiffness of the drag-tag in order to
to increase $\alpha_\text{L}$ since this implies changing the chemistry of the
label. However, the value of $\alpha_\text{L}$ is a relative measure of the
stiffness of the two components of the hybrid polyelectrolyte-label molecule (see
Eq.~\ref{eq:elfse-alphamccormick}). Therefore, an increase of $\alpha_\text{L}$
can also be achieved if we reduce the stiffness of the polyelectrolyte. The
persistence length of a polyelectrolyte can be reduced very effectively by
increasing the ionic strength of the buffer: this increases the screening of the
electrostatic repulsive interactions that stiffen the backbone of the charged
polymer. In this subsection, we investigate the effect of changing the
concentration of salt from $c_\text{S}=16 mM$ to $c_\text{S} = 1M$, which reduces
the Debye length from $\lambda_\text{D} \approx 4.2$ to $\lambda_\text{D} \approx
0.65$.

The increased electrostatic screening reduces the extension of the
polyelectrolyte chain and reduces the contribution of electrostatics to the
persistence length, which is determined to be
\begin{displaymath}
l_\text{p,PE} = 3.8 \pm 0.2, 
\end{displaymath}
whereas the label persistence is unaffected. Thus, one predicts a
hydrodynamic drag coefficient of
\begin{displaymath}
\alpha_\text{L} = \left( 0.50 \pm 0.04 \right) L,
\end{displaymath}
using Eq.~\ref{eq:elfse-alphamccormick}, instead of 
\[
\alpha_\text{L} = \left( 0.37\pm 0.03\right) L
\]
for the previous salt concentration (see Eq.~\ref{eq:alphalinear}).

\begin{figure}[htp]
\begin{center}
  \includegraphics[width=\textwidth]{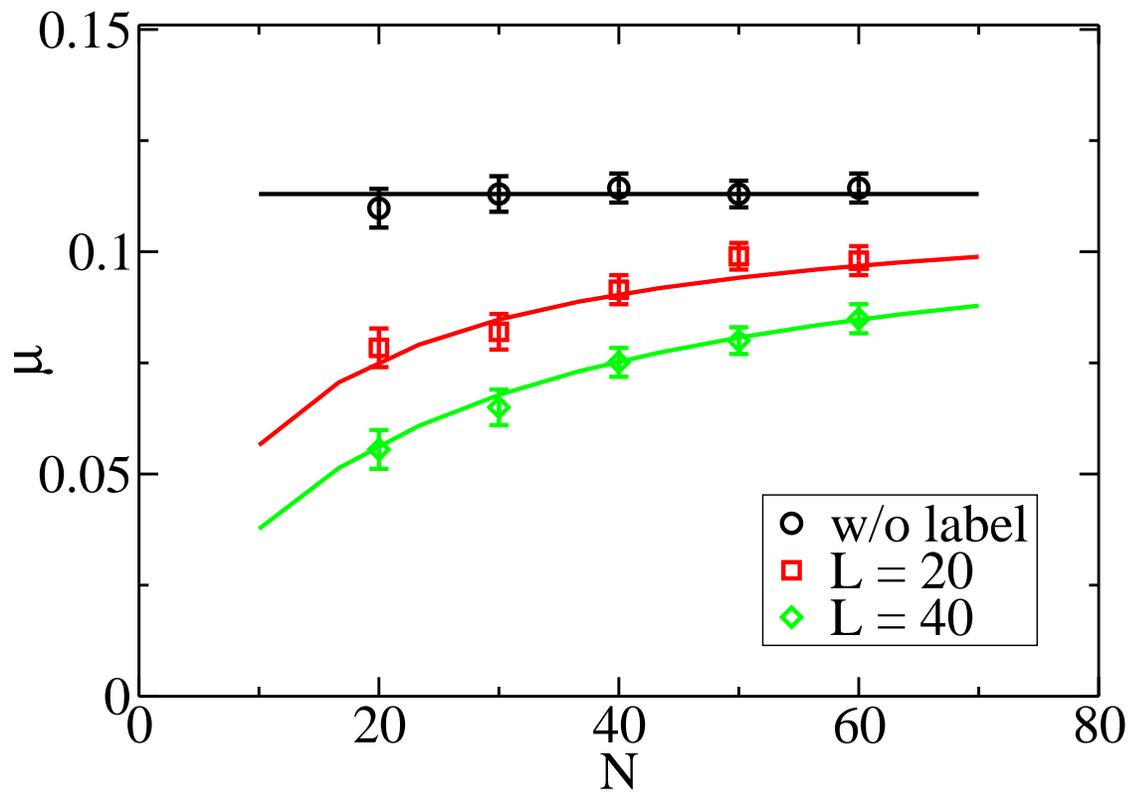}
  \caption{In the presence of 1 Mol additional salt, the persistence length
  of polyelectrolyte is reduced, changing the
  relative hydrodynamic drag $\alpha_\text{L}$ of the label. The observed
  mobility of the polyelectrolyte molecules for two linear labels of
  length 20 and 40 are compared to the prediction using Equation~\ref{eq:elfse-alphamccormick}.}
  \label{fig:elfse-highsalt}
\end{center}
\end{figure}

The change in electrophoretic mobilities for labels of lengths $L=20$ and
$L=40$ can be seen in Figure~\ref{fig:elfse-highsalt}. Please note that the
free-draining mobility $\mu_0$ also changes due to the fact that the additional
salt also increases the screening of the polyelectrolyte charge, thus reducing
the net force from the external field: 
\begin{displaymath}
 \mu_0 = 0.113 \pm 0.002. 
\end{displaymath}

The reduction in the absolute mobility $\mu_0$ together with the increase of the
diffusion coefficient due to the more compact conformation of the molecules at
higher salt concentrations negatively affect the size-selectivity as can be seen
when evaluating the resolution factor $R$ as defined by
McCormick~\etal~\cite{mccormick01a}. If we keep the polyelectrolyte length $N$,
the label length $L$, the electric field strength $E$ and the elution distance
constant then one obtains for a given $\alpha_\text{L}$ a resolution factor $R$
that only depends on the diffusion coefficient $D_0$ and the free-draining
mobility $\mu_0$:
\[
  R \sim \sqrt{D_0/\mu_0}.
\]
The way the resolution factor is defined a higher value indicates a lower
size-selectivity and thus the size-selectivity is decreased if the relative drag
of the label is increased by increasing the salt concentration. Consequently, an
increased hydrodynamic drag coefficient is less effective when achieved by adding
additional salt.

\section{Branched drag tags}\label{sec:branchedlabel}

In this section, we will investigate the use of branched labels as a possible way
to synthesize more efficient ELFSE drag-tags. First, the results obtained in a
recent experimental study by Haynes \etal are briefly reviewed~\cite{haynes05a}.
The study compared a linear polypeptide drag-tag with 30 repeat units to two
branched drag-tags, each with 5 side-chains spaced evenly along a 30 unit-long
backbone. The two different branched labels had 4 and 8 monomer long side-chains.
The drag coefficients $\alpha_\text{L}$ were obtained by measuring the mobility
of two different DNA fragments of 20 and 30 bases in length. It was found that
the value of $\alpha_\text{L}$ increases roughly linearly with the
total molecular weight of the branched label.

This astonishing observation was theoretically analysed by Nedelcu
\etal~\cite{nedelcu05a}. It was shown that the drag
coefficient is directly related to the hydrodynamic radius (as one would expect
from the blob picture), and that the linear dependence on molecular weight is
only approximately true in the limit of short side chains.

As a matter of fact, the drag provided by a linear label is always higher than
that provided by a branched label of the same molecular weight. The reason for this is
that, with a fixed length backbone, a branched polymer is essentially a compact
star polymer with a smaller hydrodynamic size than the linear equivalent. Indeed,
as the number of arms increases, the branched polymer becomes even more compact
and less favorable for ELFSE.

Based on the observations, the following optimal design using branched polymeric
labels for ELFSE was proposed: I) side chains with length comparable to the
distance between branching points, or II) two long branches located near the ends
of the molecule's backbone.

Here, the focus will be on investigating the effect of the length of the side
chains for a polymeric drag-tag with a fixed backbone length. Similar to the
structure of the label used by Haynes \etal, the label has a backbone of $L=30$
monomers to which 5 side chains are attached evenly spaced along the backbone.
The side-chain length is varied from 2 to 8 monomers, so that the total number
of monomers in the label ranges from 40 to 70. The drag coefficient of
the labels is determined by measuring the electrophoretic mobility of
polyelectrolyte chains from $N=20$ to $N=60$.

\begin{figure}[htp]
\begin{center}
  \includegraphics[width=\textwidth]{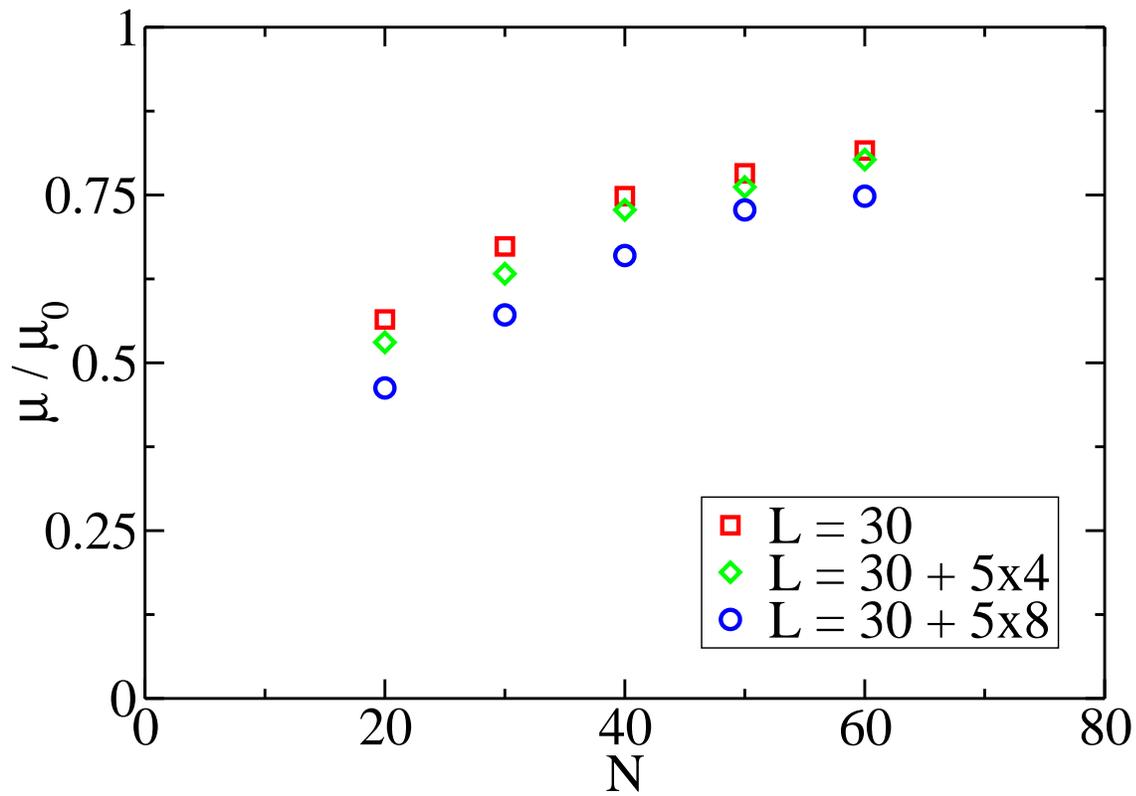}
  \caption{The reduced mobility $\mu/\mu_0$ for polyelectrolytes with an
  attached linear label with five side chains of length 4 and 8 shows a more
  pronounced slowdown than for the label without side chains.}
  \label{fig:elfse-branchedmum0}
\end{center} 
\end{figure}

Figure~\ref{fig:elfse-branchedmum0} shows the simulation results for a 30
monomeric label without side chains and with the tetra and octamer side
chains. To analyze the hydrodynamic drag of the branched labels in
detail, $\alpha_\text{L}$ is determined according to
Equation~\ref{eq:elfse-alphal}. The obtained $\alpha_\text{L}$ values are
compared to the corresponding value of a purely linear drag-tag with the same
number of monomers.

\begin{figure}[htp]
\begin{center}
  \includegraphics[width=\textwidth]{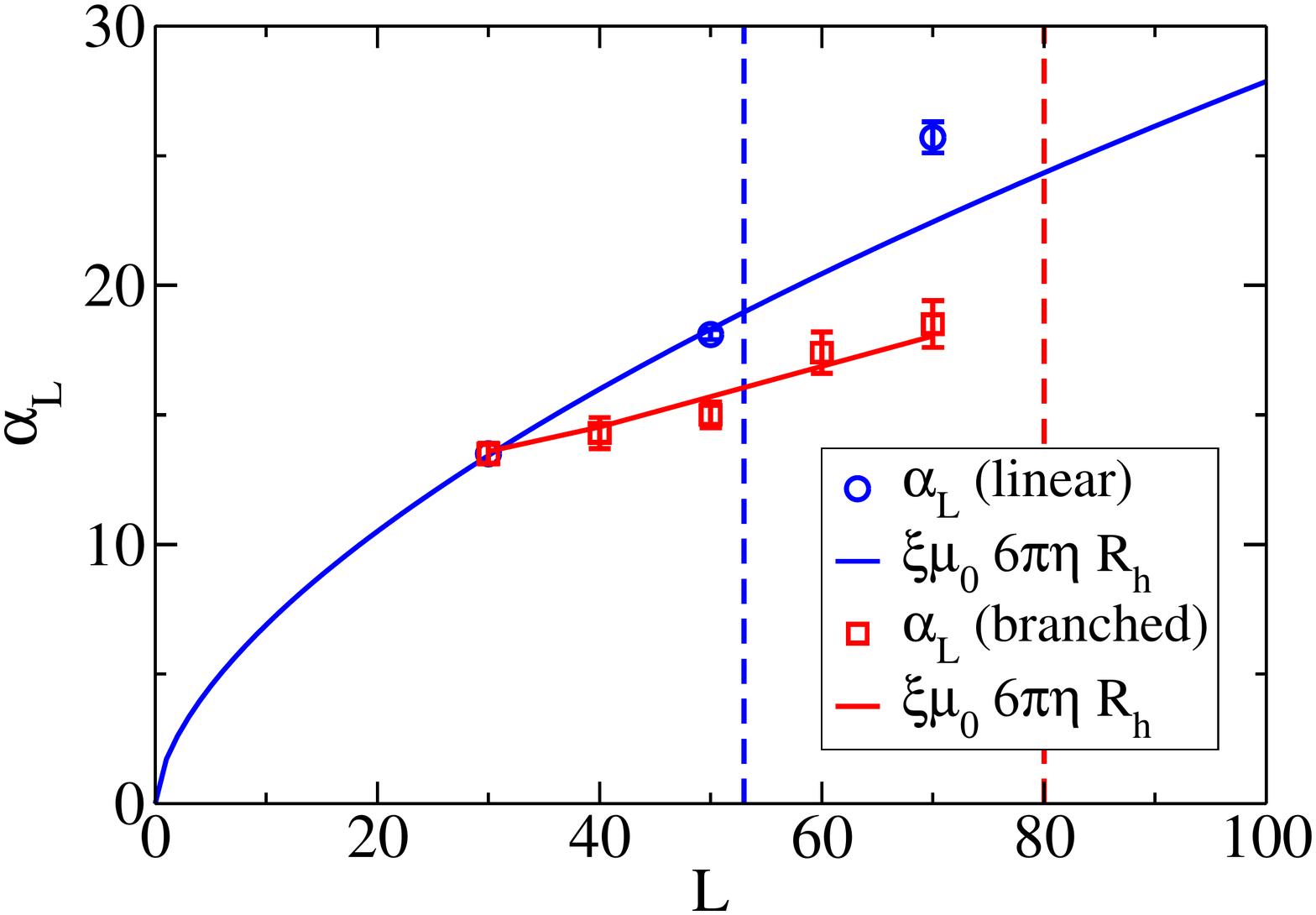}
  \caption{The hydrodynamic drag coefficient $\alpha_\text{L}$ of a branched
  polymeric label is compared to the previously determined drag of a
  linear label. $L$ is the total number of monomers. As long as the
  hydrodynamic radius $R_\text{h}$ of the label is smaller than the Debye
  length $\lambda_\text{D}$, the $\alpha_\text{L}$ is given by
  Equation~\ref{eq:elfse-alpharh}. The vertical lines indicate the number of
  monomers $L$ for which $R_\text{h}(L) \approx \lambda_\text{D}$ obtained
  from simulations.}
  \label{fig:elfse-branchedalpha}
\end{center}
\end{figure}

Figure~\ref{fig:elfse-branchedalpha} confirms the work by Nedelcu, showing that
the label with the highest drag per monomer is the linear label. For the same
number of monomers $L$, the hydrodynamic drag coefficient $\alpha_\text{L}$ of
the linear label is higher than that of the branched one. But it also shows that
the addition of side chains can be used to increase the hydrodynamic drag of the
label. This is attributed to two effects: firstly, the hydrodynamic size of the
label is increased as the side chains extend from the label. Of similar
importance is the second effect, namely that the side chains stiffen the label
due to steric repulsion with the backbone, increasing the overall persistence
length and increasing the linear length of the backbone. In fact, this is the
main contribution to the increase for the side chains of two and four monomers
as we confirmed by measuring the change in persistence length of the backbone.

Interestingly, the drag coefficients obtained for the labels show a scaling with
the hydrodynamic radius $R_\text{h}$, as given by
Equation~\ref{eq:elfse-alpharh}. Since the polymer coil formed by the
branched label is more compact, it is less penetrated by ions and,
therefore, the prediction of Equation~\ref{eq:elfse-alpharh} remains
valid for a higher number of monomers compared to the linear label.

The experimentally observed linear scaling with $L$ can be attributed to
seemingly linear relationship between $R_\text{h}$ and $L$, but, as
Nedelcu \etal have shown before,  this is only approximately true
in the case of side chains smaller or equal to the spacing along the
backbone. The only relevant quantity in all cases is the hydrodynamic radius
and its contribution to the hydrodynamic drag, as formulated in
Equation~\ref{eq:elfse-alpharh}.

Although linear labels remain preferable as long as the pure hydrodynamic drag
coefficient $\alpha_\text{L}$ per molecular weight is concerned, branched
polymers offer practical advantages because of the possibility of synthesizing
larger and somewhat stiffer monodisperse molecules in a simple, stepwise way.

\section{Micellar drag tags}\label{sec:micellelabel}

Recently, Grosser \etal~\cite{grosser07a,savard08a} proposed another promising
class of drag-tags that in principle can provide very large hydrodynamic drag
coefficients $\alpha_\text{L}$. They used nonionic surfactant Triton X-100
micelles that attach to PNAA-taged (PNA amphiphile) DNA strands. The micelles
are water-soluble and are created and destroyed on a timescale of milliseconds to seconds, forming
a fairly monodisperse populations of structures with a tunable size and
morphology. During the whole electrophoresis time, a single DNA strand attaches
to a large number of different micelles. Of importance for the ELFSE application
is the fact that this leads to an averaging effect between micelles of different
sizes for the individual DNA strand, meaning the DNA can be though of as having
a drag tag of fixed size $\langle R \rangle$, where $\langle R \rangle$ is the
average micelle size. Only with this averaging, the natural polydispersity of the
micelles is overcome and a measurement with a size resolution up to a couple of
base pairs is possible.

As a free DNA strand quickly attaches to a new micelle, the DNA is bound to a
micelle most of the electrophoresis time. Consequently, the
transiently bound micelles provide about the same hydrodynamic drag as a covalently bound drag-tag
of similar size would provide. The reported $\alpha_\text{L}$ values range
between 33 and 58 for a single micelle, depending on the micelle type and the PNAA molecule used
for connecting to the DNA strand. Savard \etal~\cite{savard08a} showed that
dual-tagging of the DNA, \ie attaching a PNAA molecule to both ends of the DNA
strand so that two micelles are transiently bound can increase the hydrodynamic
drag even further.

In this study, four different micelles with radius $R=2$ to $R=5$ are attached to
polyelectrolyte chains of different lengths. Neither the attaching and detaching
process nor the forming of the micelles themselves are modelled explicitly;
only the hydrodynamic drag of a covalently bound spherical drag-tag is
investigated. The results by Grosser and Savard show that the polyelectrolyte is
in fact attached to micelles for most of the time and only spends a small
fraction of time without a drag-tag. As the micelle size can be exactly chosen
in simulations, the averaging procedure resulting from the attaching and detaching
process does not need to be included.

\begin{figure}[htp]
\begin{center}
  \includegraphics[width=\textwidth]{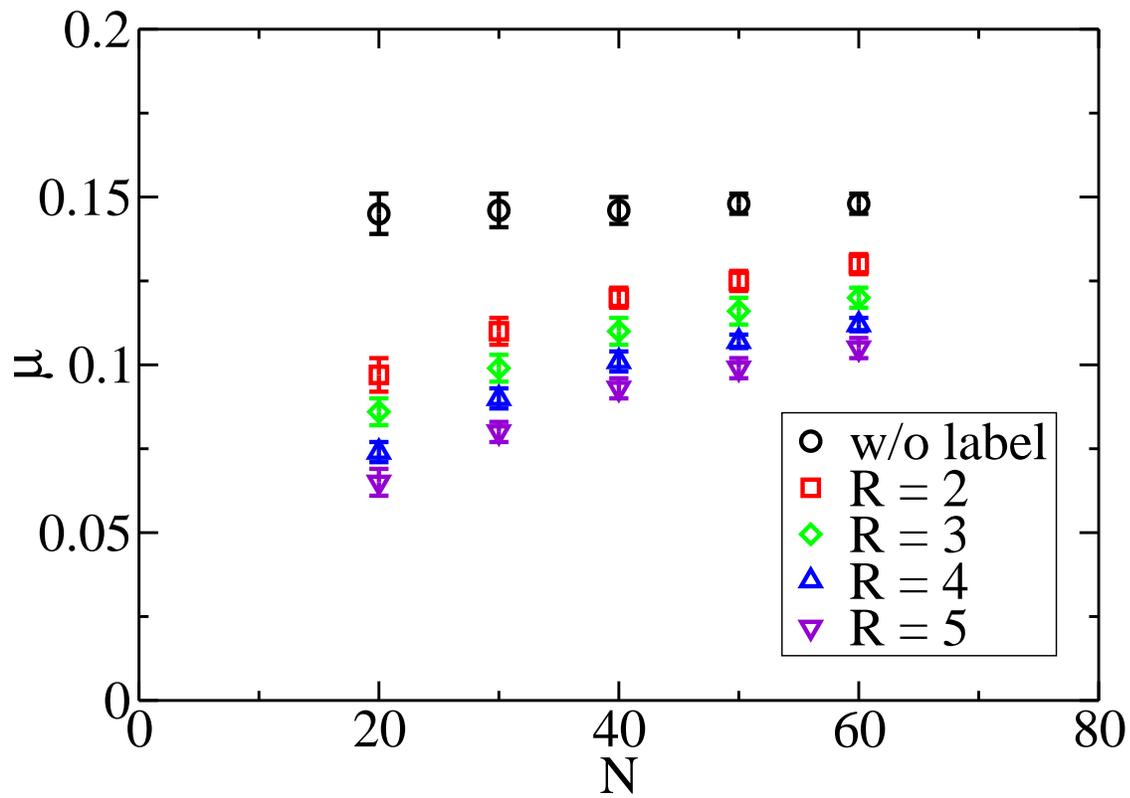}
  \caption{The electrophoretic mobility $\mu$ as a function of the
  polyelectrolyte length $N$ becomes size dependent when a micellar drag-tag
  is attached. The magnitude of the slowdown depends on the radius $R$ of
  the micelle.}
  \label{fig:elfse-micellemob}
\end{center}
\end{figure}

\begin{figure}[htp]
\begin{center}
  \includegraphics[width=\textwidth]{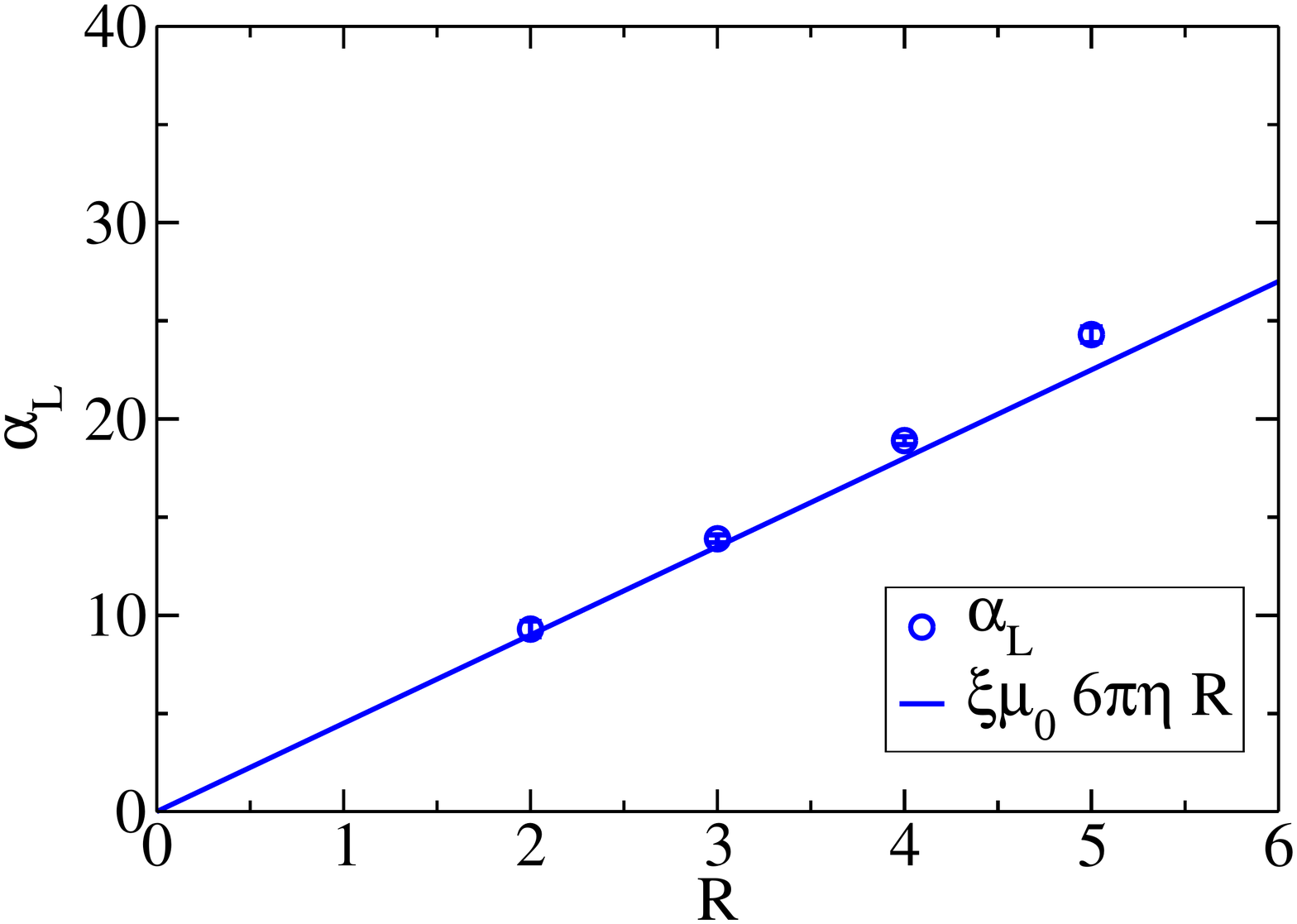}
  \caption{The effective hydrodynamic drag coefficient $\alpha_\text{L}$ of a
  micellar drag-tag is directly proportional to its radius $R$. 
  Equation~\ref{eq:elfse-alpharh} (the solid line) gives a very good prediction
  of the drag coefficient for all tested micelles.}
\label{fig:elfse-micellealpha}
\end{center}
\end{figure}

Figure~\ref{fig:elfse-micellemob} shows that micellar drag-tags can be
successfully used for elec\-tro\-pho\-retic separation of polyelectrolyte
chains. The values for $\alpha_\text{L}$ are obtained as before and compared to
Equation~\ref{eq:elfse-alpharh}, which correctly predicts the observed behaviour.
With the chosen micelle radius of $R=5$, drag coefficients up to
$\alpha_\text{L} = 24.3$ are achieved.

The results show that the hydrodynamic drag is directly related to the size
of the micelle, as can be seen in Figure~\ref{fig:elfse-micellealpha}. A linear
increase with the radius is observed, as expected from Stokes theory. Again,
Equation~\ref{eq:elfse-alpharh} correctly predicts the drag coefficient,
clearly indicating that only the hydrodynamic size of the drag-tag is
important, not the number of units it is made of or the weight associated with it.

The authors strongly believe that the use of micellar drag-tags has great
potential for the further advancement of end-labeled free-solution
electrophoresis. Especially the tunable size makes them ideal candidates, as the
drag coefficient can be optimised to the lengths of polyelectrolyte fragments to
be analysed. Currently, we are investigating the usability of cationic micelles
as drag-tags carrying positive charges, which create an additional force on the
polyelectrolyte-label compound, possibly enhancing the size separation.

\section{Conclusion}

In this paper, we have presented a detailed study of end-labeled free-solution
electrophoresis (using various hydrodynamic drag-tags) by coarse-grained
molecular dynamics simulations. Linear, branched and micellar drag-tags were investigated.
The simulations support the theoretical predictions and can be matched
quantitatively to it. This enables the use of computer simulation as a tool to
support the design of improved hydrodynamic drag-tags usable for electrophoretic
separation of polyelectrolytes in free-solution.

It was specifically shown that the drag coefficient of the label is
determined by its hydrodynamic size and not by its weight. The hydrodynamic drag
per label monomer is largest for linear labels, but experimental restrictions in the synthesis of
such labels and the monodispersity requirement limit their
practical applicability.

The addition of side chains to a linear label offers the possibility to increase
the hydrodynamic size without having to increase the linear length of the label.
The synthesis process creates perfectly monodisperse labels. It was shown that
the label efficiency is increased with the length of the side chains for the
drag-tag sizes studied in this work. In addition to increasing the lateral size
of the drag-tag, the side chains also increase the persistence of the backbone
and thus contribute in two different ways to the increased hydrodynamic size.
Especially the steric stabilisation of the linear backbone is responsible for an
initial increase of the drag-coefficient with the total number of monomers of the
label, \ie with the molecular weight. For longer side chains, the lateral
contribution to the hydrodynamic radius becomes more important.

The third class of labels investigated seems very promising for the future of
ELFSE. Transiently bound micelles provide a significantly higher hydrodynamic
drag, as they can be prepared with a large hydrodynamic radius. Additionally,
the time averaging by attaching to many different micelles over the
electrophoresis time span helps to meet the monodispersity criteria. This
study showed that the hydrodynamic drag is directly proportional to the
hydrodynamic radius of the micelle. The efficiency of this method is, in
principle, only limited by the size of labels that can be prepared.

Our results demonstrate convincingly that theory and computer models can support
the experimental progress towards the design of novel improved drag-tags, thereby
extending the applicability of the ELFSE technique. The usability of charged
drag-tags is currently under investigation. 

\begin{acknowledgement}
Funds from the the Volkswagen foundation, and the DAAD are gratefully
acknowledged. All simulations were carried out on the compute cluster of the Center for Scientic computing
at Goethe University Frankfurt. GWS would like to acknowledge the support from
the Natural Science and Engineering Research Council of Canada.
\end{acknowledgement}

\bibliographystyle{achemso}
\bibliography{simbio} 

\end{document}